\documentclass{PoS}
\usepackage{amsmath}

\DeclareMathOperator*{\argmin}{arg\,min}
\usepackage{amsfonts}
\usepackage{amssymb}
\usepackage{times}
\usepackage{fontenc}
\usepackage{mathptmx}
\usepackage{graphicx}
\usepackage{subfig}
\usepackage{color}
\usepackage{lineno}

\title{Calibration and first measurements of \textit{MuTe}: a hybrid Muon Telescope for geological structures}

\ShortTitle{Calibration and first measurements of MuTe..}

\author{\speaker{Jes\'us Pe\~na-Rodr\'iguez}$^{a}$, Adriana V\'asquez-Ram\'irez$^{a}$, Jos\'e D. Sanabria-G\'omez$^{a}$, Luis A. N\'u\~nez$^{a,b}$, David Sierra-Porta$^{a,c}$, and Hern\'an Asorey$^{d,e}$\\
        \llap{$^{a}$} Escuela de F\'isica, Universidad Industrial de Santander, Bucaramanga-Colombia\\
        \llap{$^{b}$} Departamento de F\'{\i}sica, Universidad de Los Andes, M\'erida-Venezuela.\\
        \llap{$^{c}$} Centro de Modelado Cient\'{\i}fico, Universidad del Zulia, Maracaibo-Venezuela.\\
        \llap{$^{d}$} Departamento F\'isica M\'edica, Centro At\'omico Bariloche, Comisi\'on Nacional de Energ\'{\i}a At\'omica, Bariloche-Argentina; \\
        \llap{$^{e}$} Instituto de Tecnolog\'{\i}as en Detecci\'on y Astropart\'{\i}culas (ITEDA), Buenos Aires-Argentina.\\
        E-mail: \email{jesus.pena@correo.uis.edu.co}, \email{adriana2168921@correo.uis.edu.co},
        \email{jsanabri@uis.edu.co},
        \email{lnunez@uis.edu.co}, \email{david.sierra1@correo.uis.edu.co}, \email{asoreyh@cab.cnea.gov.ar}
        }

\abstract{In this work, we describe the calibration and first measurements in the commissioning of MuTe, a  hybrid Muon Telescope with two subdetectors --a scintillator hodoscope and a Water Cherenkov Detector (WCD)--  for imaging the inner structures of Colombian volcanoes. The hodoscope estimates the trajectories of particles impinging on the front and rear panel, while the WCD acts as a calorimeter for the through going charged particle. 

MuTe combines particle identification techniques so as to discriminate noise background from data. It filters the primary noise sources for muography, i.e., the EM-component ($e^{\pm}$) of Extensive Air Showers (EAS) and scattered/upward-coming muons. The WCD identifies Electrons/positrons events by their deposited energy identifies, while scattered and backward muons are rejected using a pico-second Time-of-Flight(ToF) system.

Muon generated events were found in the deposited energy deposited range of ($144$MeV$ < E_d <$ 400MeV), represent only about the 40$\%$ of the WCD-hodoscope acquired events. The other 60$\%$ of data is composed by ($e^{\pm}$) events under 144 MeV and multiparticle events above 400 MeV. Subsequently, low-momentum muons ($<$ 1 GeV/c), which are scattered by the volcano surface, measures a ToF $>$ 3.3 ns for traversing one meter length.

 }

\FullConference{36th International Cosmic Ray Conference -ICRC2019-\\
		July 24th - August 1st, 2019\\
		Madison, WI, U.S.A.}

\begin{document}
\section{Introduction}
Muography is a non-invasive technique --spatial resolution in the order of tens of meters-- for imaging anthropic and geologic structures implemented several decades ago\cite{George1955,AlvarezEtal1970}. Recently it has been boosted with several new successful applications such as: detection of hidden materials in containers \cite{Blanpied2015}, archaeological building scanning \cite{Morishima2017, GomezEtal2016}, nuclear plant inspection \cite{Fujii2013}, nuclear waste monitoring, underground cavities \cite{Saracino2017}, overburden of railway tunnels\cite{ThompsonEtal2019} and volcanology (\cite{TanakaOlah2019} and references therein). For sometime now, research groups in Colombia have been exploring this technique which measures the variation in the atmospheric muon flux crossing geological edifices \cite{AsoreyEtal2017B, SierraPortaEtal2018, PenaRodriguezEtal2018, GuerreroEtal2019, ParraAvila2019}. 

The flux variance between trajectories allows us to extract information about the inner density distribution of the scanned object. However, muography is affected by a density sub-estimation due to the recording of false-positive events, generated by three main phenomena: horizontal upward/backward-coming muons, low energy muons (< $1$GeV) scattered from the volcano surface \cite{Nishiyama2014B,Gomez2017} and the EM-component ($e^-$, $e^+$) of EAS.

Various methods have been tried in different experiments for the removal of this background using:  Time-of-Flight, ToF, systems for upward-coming muons rejection \cite{Marteau2014, Cimmino2017}, installation of absorbent panels for low energy muons filtering and increasing the number of sensitive panels for decreasing the probability of EAS-generated events \cite{Lesparre2012}.

Implementation of extra (absorbent or sensitive) panels affects the detector complexity and cost but, ToF systems arise as a reliable solution, filtering low energy muons with a discrimination resolution $\sim 10$MeV, with time differences of about $10$ps. 

In this paper we report the calibration method and first measurements of a Colombian hybrid Muon Telescope, MuTe, highlighting a new method for removing the background noise based on a ToF system which filters upward-coming \& low energy muons and, also the inclusion of a WCD --based on the Latin American Giant Observatory, LAGO experiences -- for the rejection of the EM-component of EAS.

\section{The instrument}

\begin{figure}[!ht]
\begin{center}
\includegraphics[width=.6\textwidth]{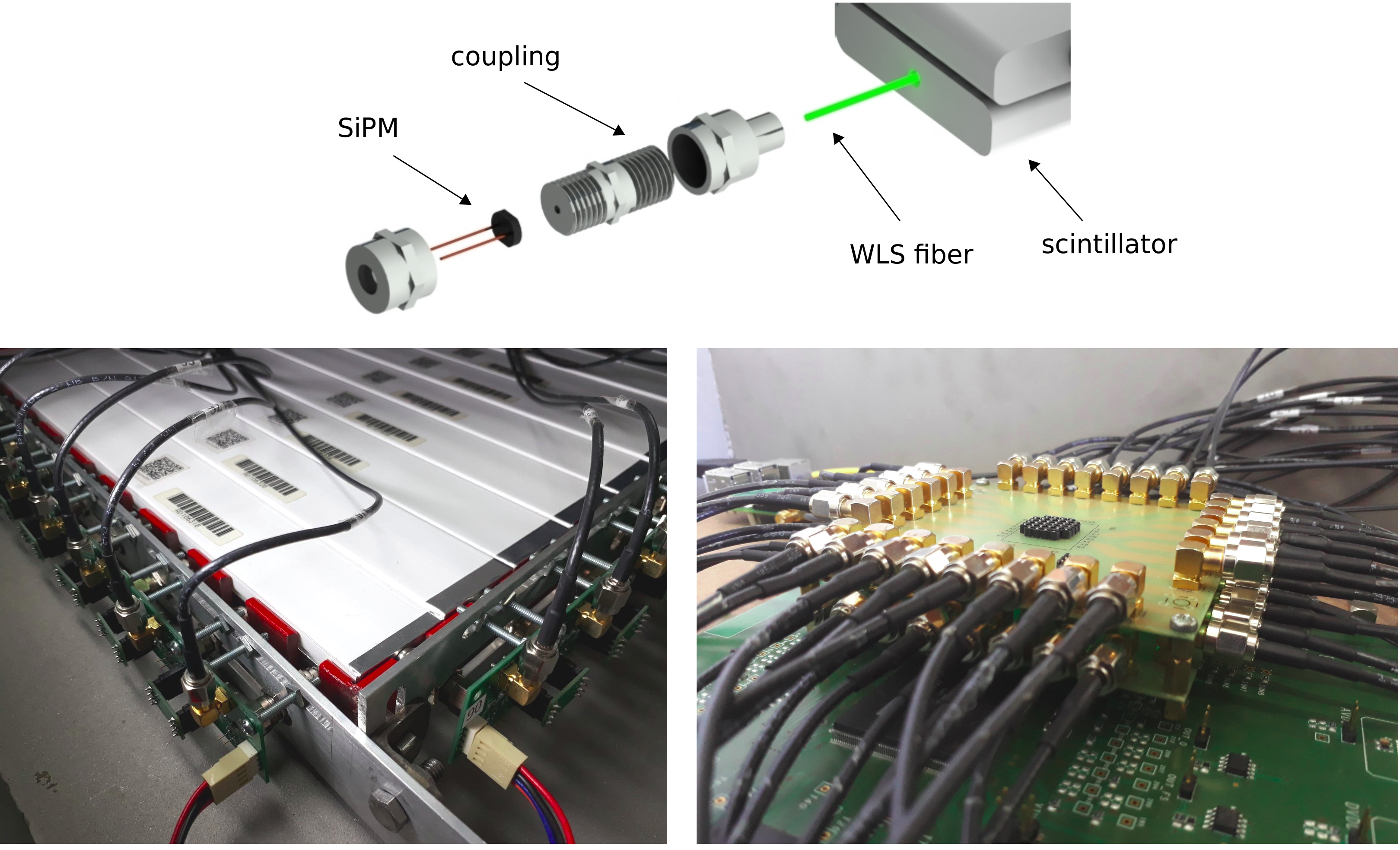}
\caption{Assembling of the Saint-Gobain WLS fiber, mechanical coupling and the Hamamatsu SiPM in a scintillator bar (up). Structure of the rear scintillator panel, the electronics front-end and the transmission lines (down-left). MAROC3A-shield connector for 60 signals coming from a scintillator panel (down-right).}
\label{Scintillator}
\end{center}
\end{figure}

MuTe is a hybrid detector with two subdetectors: a scintillator hodoscope and a WCD. The hodoscope --with an angular resolution of $26.18 \textrm{ mrad}$ for an inter-panel distance of $2.5 \textrm { m}$-- consists of two panels each of $30 \times 30$ strips of $120 \textrm{ cm} \times 4 \textrm{ cm} \times 1 \textrm{ cm}$, of polystyrene (Dow Styron $663$) with an external coating of TiO$_2$ and  dopants  ($1\%$  PPO,  $0.03 \%$ POPOP),  an absorption cut off  $\sim 40$ nm and an emission maximum at $420$ nm. 

Each strip has  a 1.8 mm hole for a wavelength shifting (WLS) multi-cladding fiber  (Saint-Gobain BCF-92 with $1.2$ mm diameter) with a core refraction index of $1.42$, an absorption peak at $410$ nm and an emission peak of $492$ nm. Each WLS fiber is coupled, with a silicon photomultiplier (SiPM, Hamamatsu S13360-1350CS). The SiPM has a photosensitive area of $1.3 \times 1.3 \textrm{ mm}^2$, $667$ pixels, a fill factor of $74\%$, a breakdown voltage of $53\pm$5 V, a gain from $10^5$ to $10^6$ and a photon-detection efficiency of $40 \%$ at $450 \textrm{nm}$.

Figure \ref{Scintillator} illustrates how each SiPM has a front-end electronics for polarisation and signal conditioning (i.e., pre-amplification for Signal-to-Noise Ratio enhancement). An ASIC MAROC3A --whose slow control parameters are set by means of a FPGA Cyclone 3-- individually amplifies and jointly discriminates the 60 signals from each panel. The data is managed by a Raspberry Pi 2 and stored in a central hard disk.

\begin{figure}[!ht]
\begin{center}
\includegraphics[width=0.55\textwidth]{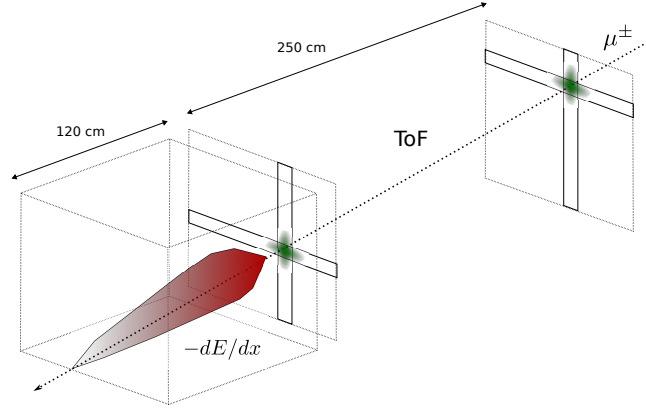}
\caption{The general sketch of the MuTe detection scheme. The hodoscope estimates the event flux per trajectory depending on the fired pixels in the frontal and the rear panel. A ToF system measures the time taken for crossing particles subtracting the nano-scale time stamp recorded in both panels. The LAGO-WCD senses the Cherenkov photons generated in the water due to the charged particle interaction. The recorded photon yield is equivalent to energy loss.}
\label{MuTeFeat}
\end{center}
\end{figure}

The cubic LAGO-WCD (120 cm side) has a Tyvek internal coating and an eight inch Hamamatsu R5912 photomultiplier tube (PMT) as the sensitive element, which provides two signals (anode and the last dynode), digitized by a $10$ bits Fast Analog to Digital Converters (FADCs) with a sampling frequency of $40$ MHz \cite{SofoHaro2016}.  This sub-detector measures (see Figure \ref{MuTeFeat}) the energy loss ($-dE/dx$) of the passing charged particles generating  Cherenkov photons and allows the muon/background discrimination, by identifying the EM-component of cosmic ray showers and low energy muons ($< 2$GeV) which are the main noise sources in muography.

A Time-to-Digital Converter (TDC) implemented on a Xilinx FPGA Spartan 6 to measure the crossing particle ToF. The TDC has two stages: a counter based on a $100$ stages delay line with a resolution of $40$ps and a course counter based on a Ring Oscillator whose resolution rises up to $5$ns. The range of the TDC system is $90$ns. The ToF is critical for filtering false-positive events related to upward-coming muons (i.e., negative ToF) and for estimating the particle momentum.

\section{Trigger system}

The MuTe trigger system has five hierarchical levels (T1, T2, T3, T4, and T5). T1 is formed by a discriminator (d1) in the MAROC3A when the scintillator pulse amplitude exceeds the threshold ($V_{th0}$), which is established by a 10 bits Digital to Analog Converter (DAC).

Level T2 is an offline condition to single out orthogonal strip coincidences, i.e., an event candidate may trigger only one x-bar and y-bar to resolve the pixel positioning. T3 trigger occurs when two events, between the frontal and the rear panel,  coincide within a time window of $7$ns (shortest panel distance $2.5$m) to 12 ns (largest panel distance $3.5m$). 

Level T4 triggers a LAGO-WCD event when a PMT dynode pulse exceeds a 100 ADC threshold ($\sim$100 mV) in the LAGO-WCD discriminator. The digitized pulse is stored in a 12 bin vector with an inter-sample time step of 25 ns. Finally, when an event candidate raises the T3 and T4 flags, the T5 level occurs, containing all the particle information: trajectory, energy loss, and ToF.

\section{Hodoscope acceptance}


The number of events detected by the hodoscope for a given direction $r_{m,n}$ ($m=i-k,n=j-l$ with $\{ i,j \}$ as the pixel coordinates for the frontal panel and $\{ k,l \}$ for the rear one) can be expressed as  $N(r_{m,n}, \Delta T)~=~I(r_{m,n}) \times \Delta T \times \mathcal{T}(r_{m,n})$ where $I$ represents the flux of events given in cm$^{-2}$sr$^{-1}$s$^{-1}$, $\Delta T$ measures the period and the acceptance is $\mathcal{T}(r_{m,n}) = S(r_{m,n}) \times \delta \Omega (r_{m,n})$; $S$ is the detection surface and $\delta \Omega$ the angular aperture. For two panels with $N_x\times N_y$ pixels we can identify $(2N_x-1)(2N_y-1)$ different discrete trajectories, where the angular aperture depends on the the inter-panel distance $D$. The MuTe hodoscope can reconstruct 3481 trajectories with a solid angle in $\theta_x=\theta_y=0$ ($r_{0,0}$) about 8$\times 10^{-4}$ sr for an inter-panel distance of $D =$134 cm. 

\section{Hodoscope and WDC calibrations}

\begin{figure}[!ht]
\begin{center}
\includegraphics[width=0.6\textwidth]{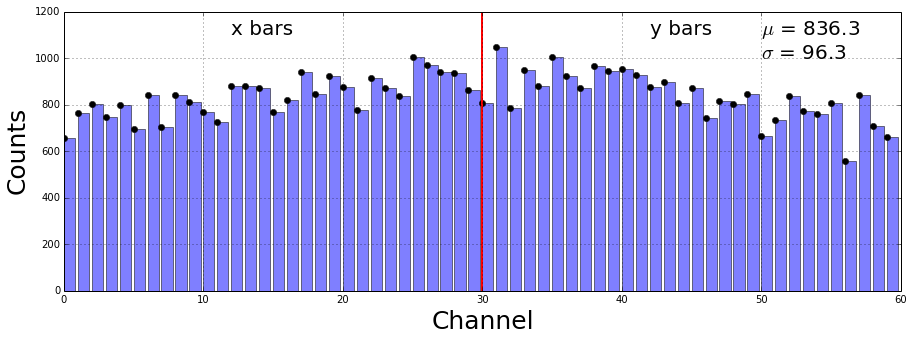}
\caption{The results of the bar equalization process in an hour of data applying the triggers T1 and T2.  The detection rate per bar is $836.3$ event/h, and the variability is $96.3$ event/h. The red line separates the $x-bar$ and $y-bar$ histograms.}
\label{Equalization}
\end{center}
\end{figure}

Several factors --material impurities during the scintillator production, WLS fiber assembling, SiPM coupling, signal conditioning, and transmission-- can cause a variable detection rate in the scintillator bars. Thus, to reduce this channel response variability, the calibration starts with an equalization stage with a weighted gain to each bar. This is done by aiming the hodoscope to 0$^{\circ}$ zenith during 1 hour. The results, illustrated in Figure \ref{Equalization} display an average detection rate --after the equalization process-- of $836.3\pm 96.3$ event/h, consistent with results reported in \cite{Lesparre2012} for about 1190 event/h. Furthermore, a bad optical coupling is exemplified in bar $Y_{26}$ because of its low rate even after equalization.

\begin{figure}[!ht]
\begin{center}
\includegraphics[width=0.5\textwidth]{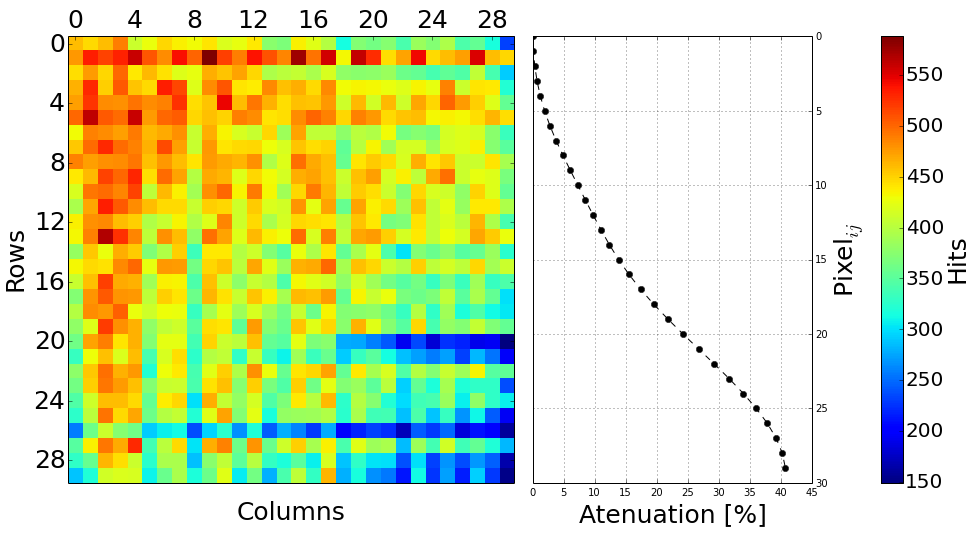}
\caption{The number of hits detected by the frontal panel for $15$ hours. The attenuation of the detection rate per pixel increases diagonally from the upper-left corner to the lower-right one. The most active pixel reaches $33.3$ event/h whereas the least active only $13$ events/h which means an attenuation about $40\%$. }
\label{Attenuation}
\end{center}
\end{figure}

The number of photons reaching the SiPM after a particle-scintillator interaction depends on the occurrence point. The higher the distance between the SiPM and the interaction point, the higher the attenuation in the photon yield. As a result of this effect, the scintillator panel is less sensitive (up to $40\%$) in the furthest corner from the SiPM placement as is shown in Figure \ref{Attenuation}.

On the other hand, the LAGO-WCD hardware calibration process consists of finding the optimum operation point of the PMT \cite{Leon2018}. Such a point, located at the middle of a region called ``plateau'', depends on the bias voltage $V_b$ and the discrimination threshold. 

\begin{figure}[!ht]
\begin{center}
\includegraphics[width=0.6\textwidth]{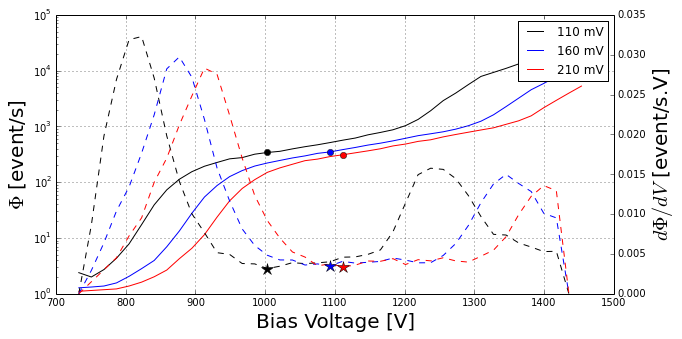}
\caption{Detected event rate ranging from 740 V to 1450 V for a threshold of 110 mV (black), 160 mV (blue) and 210 mV (red). The optimum point for the LAGO-WCD functioning for each case is located at the star points on the plateau regions.}
\label{WCDOpt}
\end{center}
\end{figure}
The bias voltage was varied from $740$ V to $1450$ V in steps of $102$ V, estimating the event rate for a 10 minutes time lapse. The optimum working point was selected using the minimization of the rate function derivative as follows
\begin{equation}
V^{\ast}_b = \argmin \frac{\textrm{d}\Phi}{\textrm{d} V} 
\end{equation}
In figure \ref{WCDOpt} the solid lines represent the LAGO-WCD detection rate for a threshold of $110$mV (black), $160$mV (blue) and $210$mV (red), while the dotted ones show their derivatives with the optimum operation voltages (shown by stars) at $1000$V($110$mV), $1096$V($160$mV), and $1109$V($210$mV).                                            

An important point is the estimation  --from the charge histogram-- of the energy loss of particles crossing the LAGO-WCD: a muon losses about 2 MeV/cm in water \cite{Asorey2012, Vasquez2018}.

\begin{figure}[!ht]
\begin{center}
\includegraphics[width=0.5\textwidth]{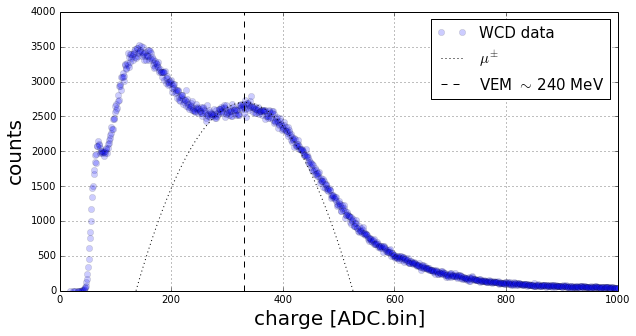}
\caption{LAGO-WCD charge histogram. The highest hump represents the electromagnetic component of the detected events, with a mean value of 140 ADC.bin. The muonic component is fitted by a quadratic distribution with 1 VEM (331 ADC.bin) mean. }
\label{Histogram}
\end{center}
\end{figure}

Figure \ref{Histogram} displays an one-hour charge histogram. It has two big humps; the first ($\sim $140 ADC.bin) due to the EM-component of EAS (i.e., electrons, positrons, and gammas) and the second one caused by the muonic component \cite{Asorey2012}. The deposited energy calibration is made taking into account the Vertical Muon Equivalent (VEM) point, which represents the deposited charge of the Cherenkov photons due to vertical muons ($\sim$0$^{\circ}$ zenith).

Subsequently, the energy loss per charge unit (ADC.bin) is defined as,
\begin{equation}
    E_{loss} =  (\text{ADC.bin})\frac{240 (\text{MeV})}{ \text{VEM}_q(\text{ADC.bin})}
\end{equation}
where (ADC.bin) is a unit of charge (i.e. equivalent to $0.72$ MeV) and VEM$_q$ is the charge at the muon hump mean ($331.4$ ADC.bin).

\section{First measurements}

\begin{figure}[h!]
\centering
\includegraphics[scale=0.22]{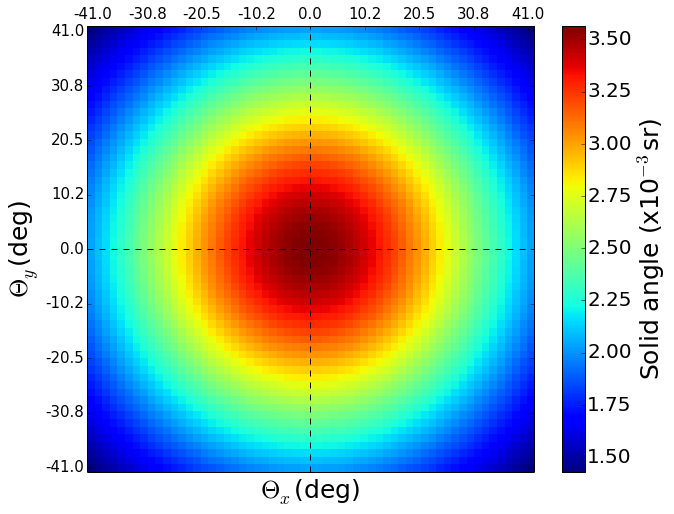} 
\includegraphics[scale=0.22]{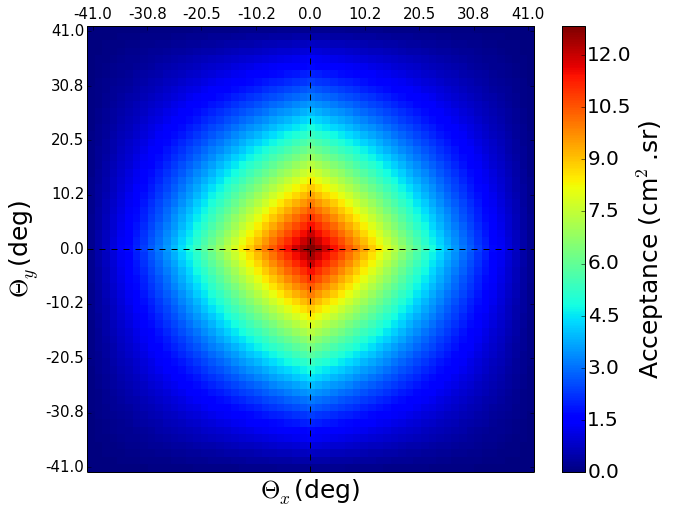}
\includegraphics[scale=0.22]{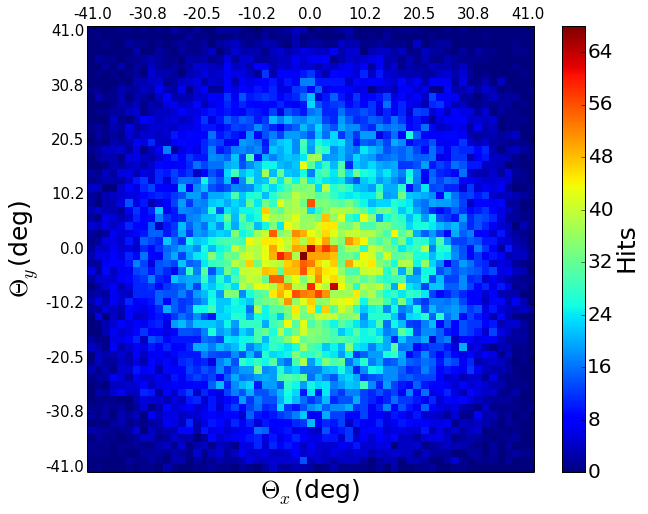}
\includegraphics[scale=0.22]{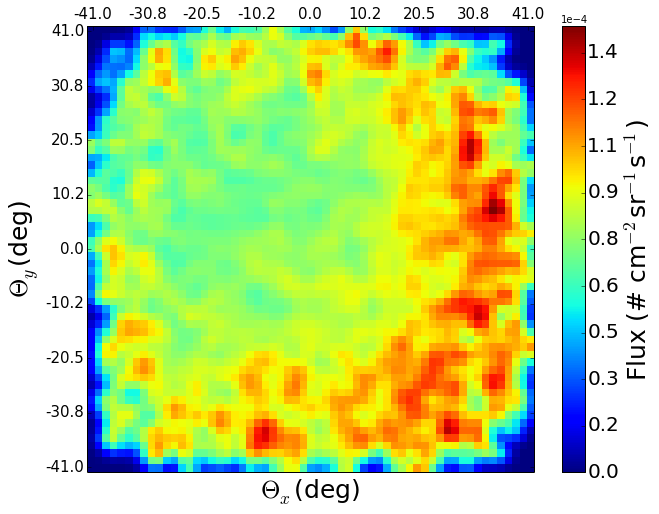}
\caption{Angular properties of the hodoscope with $N_x=N_y =$ 30, $d =$ 4 cm y $D =$ 134cm. Solid angle histogram as a function of the traversed trajectory (top-left). Hodoscope acceptance histogram (top-right). Number of events recorded by the hodoscope during 15 hours (bottom-left). Detected flux taking into account an ideal acceptance (bottom-right).}
\label{Flux}
\end{figure}

After the calibration process, the MuTe hodoscope acquired data for 15 hours. Figure \ref{Flux} displays the number of hits and the flux depending on the angular coordinates, inside the laboratory under a concrete shielding of about 30 cm. As expected, the number of detected events for orthogonal directions ($r_{0,0}$) is greater than for others, reaching 68 hits in 15 hours. The flux has an annular modulation due to a dome shaped building placed above the MuTe. The maximum flux is about 1.4 $\times  10^{-4}$ cm$^{-2}$sr$^{-1}$s$^{-1}$.

\begin{figure}[!ht]
\begin{center}
\includegraphics[width=0.6\textwidth]{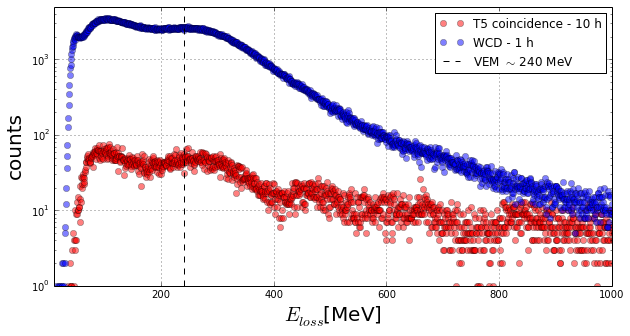}
\caption{Deposited energy histogram for the T4 (blue) and T5 events (red) detected in the LAGO-WCD.}
\label{Rejection}
\end{center}
\end{figure}

 Figure \ref{Rejection} shows the charge histogram for all the events detected during 1 hour by the LAGO-WCD (blue) and for the events that triggered T5(red) for a period of 10 hours. The T5 events represent only about $0.2\%$ of the total number of events recorded by the whole LAGO-WCD. The low T5 flux is  due to the diminished CR flux 1$\%$ at a zenith angle of $\> 70^{\circ}$ compared with $0^{\circ}$ zenith. To test this background rejection capability of the LAGO-WCD, MuTe was pointed to $90^{\circ}$ zenith with D = 250 cm.

On the other hand, the muonic component of the T5 events, located around the VEM with a standard deviation of $2\sigma$, represents about the $40\%$ of the T5 data. The background is composed of two regions: below $144$ MeV is made of by the EM particles component and above 400 MeV  dominated by the multiple-particle events.

\section{Conclusions}

This work reports the calibration and first measurements of a hybrid muon detector capable of rejecting the background signals in muography generated by the soft-component of EAS, low-momentum and upward-coming muons.

 Muon generated events deposited in a energy range of ($144$MeV$ < dE/dx <$ 400MeV), represent only about the 40$\%$ of the WCD-hodoscope acquired events. The other 60$\%$ of data is composed of ($e^{\pm}$) events below $144$MeV and multiparticle events above $400$MeV. Subsequently, low-momentum muons ($<$ 1 GeV/c), which are scattered by the volcano surface, are measured with a velocity of $>$ 0.3 m/ns.

Several reports show that background noise in muography --due to electrons, positron, and low-momentum muons--  could rise as high as $60\%$ of the acquired data \cite{Nishiyama2014B,Gomez2017}.

The Colombia MuTe combines a conventional muography method with particle identification techniques to discriminate noise background from data. This approach opens a window towards a new kind of hybrid muon telescopes with a better understanding of the background-origin and the phenomena behind the muography technique.

\acknowledgments
The authors acknowledge the financial support of  Departamento Administrativo de Ciencia, Tecnolog\'ia e Innovaci\'on of Colombia (ColCiencias) under contract FP44842-082-2015 and to the Programa de Cooperaci\'on Nivel II (PCB-II) MINCYT-CONICET-COLCIENCIAS 2015, under project CO/15/02. We are particularly thankful to the Latin American Giant Observatory Collaboration and to Pierre Auger Observatory for their permanent support and inspirations.



\end{document}